\documentclass[aps,prl,twocolumn,showpacs,floatfix,superscriptaddress]{revtex4}
\usepackage{dcolumn}
\usepackage{bm}
\usepackage{amsmath,amssymb,graphicx}

\begin{document}
\preprint{Submitted to Phys. Rev. Lett.}

\title{Resonant Raman Spectroscopy of Armchair Carbon Nanotubes: \\Absence of ``Metallic'' Signature}

\author{E.~H.~H\'{a}roz}
\affiliation{Department of Electrical and Computer Engineering, Rice University, Houston, Texas 77005, USA}
%\affiliation{Department of Physics and Astronomy, Rice University, Houston, Texas 77005, USA}
\affiliation{The Richard E. Smalley Institute for Nanoscale Science and Technology, Rice University, Houston, Texas 77005, USA}

\author{J.~G.~Duque}
\affiliation{Chemistry Division, Los Alamos National Laboratory, Los Alamos, New Mexico 87545, USA}
\affiliation{Center for Integrated Nanotechnologies, Los Alamos National Laboratory, Los Alamos, New Mexico 87545, USA}

\author{W.~D.~Rice}
\affiliation{Department of Electrical and Computer Engineering, Rice University, Houston, Texas 77005, USA}
\affiliation{The Richard E. Smalley Institute for Nanoscale Science and Technology, Rice University, Houston, Texas 77005, USA}
\affiliation{Department of Physics and Astronomy, Rice University, Houston, Texas 77005, USA}

\author{C.~G.~Densmore}
\affiliation{Chemistry Division, Los Alamos National Laboratory, Los Alamos, New Mexico 87545, USA}

\author{J.~Kono}
\email[]{kono@rice.edu}
%\homepage[]{www.ece.rice.edu/~kono}
\thanks{corresponding author.}
\affiliation{Department of Electrical and Computer Engineering, Rice University, Houston, Texas 77005, USA}
\affiliation{The Richard E. Smalley Institute for Nanoscale Science and Technology, Rice University, Houston, Texas 77005, USA}
\affiliation{Department of Physics and Astronomy, Rice University, Houston, Texas 77005, USA}

\author{S.~K.~Doorn}
\email[]{skdoorn@lanl.gov}
%\homepage[]{www.ece.rice.edu/~kono}
\thanks{corresponding author.}
\affiliation{Center for Integrated Nanotechnologies, Los Alamos National Laboratory, Los Alamos, New Mexico 87545, USA}

\date{\today}
%%%%%%%%%%%%%%%%%%%%%%%%%%%%%%%%%%%%%%%%%%%%%%%%
\begin{abstract}
The appearance of a broad peak at $\sim$1550~cm$^{-1}$ (G$^{-}$ peak) in carbon nanotube resonant Raman scattering spectra has been conventionally attributed to the presence of metallic nanotubes.  Here, we present resonant Raman measurements on macroscopic nanotube ensembles enriched in armchair species prepared through density gradient ultracentrifugation.  Our data clearly demonstrate that the broad G$^{-}$ mode is absent for armchair structures and appears only when the excitation laser is resonant with non-armchair ``metals.''  Due to the large number ($\sim$10$^{10}$) of nanotubes across several armchair species probed, our work firmly establishes a general correlation between G-band lineshape and nanotube structure.
\end{abstract}

\maketitle

%%%%% Introduction %%%%%

Armchair carbon nanotubes have a distinguished status among all the members, or species, of the single-walled carbon nanotube (SWNT) family.  Being the only truly gapless species in the family, they are expected to exhibit some of the unusual properties characteristic of one-dimensional (1-D) metals~\cite{Giamarchi04Book}.  While strong electron-electron interactions can lead to the formation of Luttinger liquid states, strong electron-phonon interactions can renormalize phonon frequencies and lifetimes (Kohn anomalies) and induce Peierls lattice instabilities, especially in small-diameter nanotubes~\cite{DubayetAl02PRL,BohnenetAl04PRL,ConnetableetAl05PRL,BarnettetAl05PRB}.  At the same time, as exceptionally conductive wires~\cite{Yaoetal00PRL}, they are promising for electronic applications such as nanocircuit inter-connects and power transmission cables.

Recently, much progress has been made~\cite{DubayetAl02PRL,LazzerietAl06PRB,IshikawaAndo06JPSJ,PiscanecetAl07PRB} in understanding electron-phonon interactions and their consequences in Raman scattering spectra in ``metallic'' SWNTs, in which the chiral indices ($n$,$m$) satisfy $\nu$ $\equiv$ $(n-m)$~mod~3~=~0, as opposed to semiconducting SWNTs, for which $\nu$~=~$\pm$1.  Note that, while armchair ($n$~=~$m$) nanotubes are metallic (gapless), non-armchair (or $n \neq m$) $\nu$~=~0 tubes have small curvature-induced bandgaps, i.e., they are in fact narrow-gap semiconductors~\cite{Hamadaetal92PRL,KaneMele97PRL}.  Of particular interest is the so-called G-band, a Raman-active optical phonon feature originating from the in-plane C-C stretching mode of $sp^{2}$-hybridized carbon.  In SWNTs, the G-band is split in two, the G$^{+}$ and G$^{-}$ peaks, due to the curvature-induced inequality of the two bond-displacement directions.  For $\nu$~=~0 tubes, the higher-frequency mode (G$^{+}$) is a narrow Lorentzian peak, while the lower-frequency mode (G$^{-}$) is extremely broad.  Earlier theoretical studies described this broad feature as a Breit-Wigner-Fano lineshape due to the coupling of phonons with an electronic continuum~\cite{BrownetAl01PRB} or low-frequency plasmons~\cite{Kempa02PRB}, but there is now accumulating consensus that the broad G$^{-}$ peak is a softened and broadened longitudinal optical (LO) phonon feature, a hallmark of Kohn anomalies~\cite{DubayetAl02PRL,LazzerietAl06PRB,IshikawaAndo06JPSJ,PiscanecetAl07PRB,NguyenetAl07PRL,WuetAl07PRL,FarhatetAl07PRL}.  Through either scenario, this broad G$^{-}$ feature has conventionally been known to be a ``metallic'' feature, indicating the presence of metallic tubes.

In this Letter, we present detailed wavelength-dependent Raman scattering measurements on a macroscopic ensemble of SWNTs enriched in armchair (or $n$ = $m$) nanotubes produced via density gradient ultracentrifugation~\cite{HarozetAl10ACS}.  Our G-band spectra clearly show that the broad G$^{-}$ mode is absent for armchair structures and only occurs for non-armchair (or $n \neq m$) ``metallic'' nanotubes.  Namely, the conventional method for identifying metallic nanotubes by observing a broad G$^-$ peak does not apply to the only truly metallic species, i.e., armchair nanotubes.  This supports an earlier conclusion based on a small number of single-tube measurements~\cite{WuetAl07PRL,MicheletAl09PRB} and negates some claims that armchair nanotubes also show a broad G$^{-}$ feature~\cite{SasakietAl08PRB,ParketAl09PRB}.  Furthermore, this result firmly establishes a general correlation between G-band lineshape and nanotube structure due to the sampling of a statistically significant number ($\sim$ 10$^{10}$) of nanotubes.

%%%%% Samples and Experimental Methods %%%%%

Density gradient ultracentrifugation (DGU), using our previously reported procedures, was employed to create armchair-enriched (AE-SWNT)~\cite{HarozetAl10ACS} and ``metal''-enriched (ME-SWNT)~\cite{NiyogietAl09JACS}, aqueous surfactant suspensions of SWNTs, synthesized by the high pressure carbon monoxide method (HiPco) at Rice University.  The ME-SWNT was a mixed ensemble containing both armchair and non-armchair (narrow-gap) $\nu$~=~0 ``metallic'' tubes.  All DGU samples were dialyzed into surfactant and water to remove the density gradient medium.  The as-produced HiPco SWNT reference sample (AP-SWNT) was produced using the standard ultracentrifugation technique in 1\% (wt./vol.) sodium deoxycholate (water)~\cite{OconnelletAl02Science,DuqueetAl10ACS}.

%This may need to be modified some after discussion with Juan regarding the LANL sample%

%Optical absorption spectroscopy was performed in the 400-1350~nm range in 1~nm steps on an ultraviolet-visible-near-infrared, double-beam spectrophotometer (Shimadzu UV-3101PC scanning spectrophotometer) through a 10~mm path length quartz cuvette using a 1\% (wt./vol.) DOC (water) reference.
Resonant CW Raman spectroscopy was performed in a backscattering geometry with frequency-doubled Ti:Sapphire laser excitation (440-500~nm), Ar$^+$ ion laser discrete lines (501.7 and 514.5~nm), and tunable dye laser excitation [with Rhodamine 6G (552-615~nm) and Kiton Red (610-680~nm)].  The laser power at the sample was maintained at 25~mW with a spot size of $\sim$300~$\mu$m.  Individual Stokes-shift spectra were obtained at 5~minute integrations using a CCD camera mounted on a triple monochromator.  The non-resonant Raman spectrum of 4-acetamidophenol, acquired at each excitation wavelength, was used for both frequency calibration and for correcting intensities for instrument response.  All data were taken at room temperature and ambient pressure.

%%%%% Experimental Results %%%%%

%%%%% Figure 1 %%%%%
\begin{figure}
\includegraphics[scale=0.53]{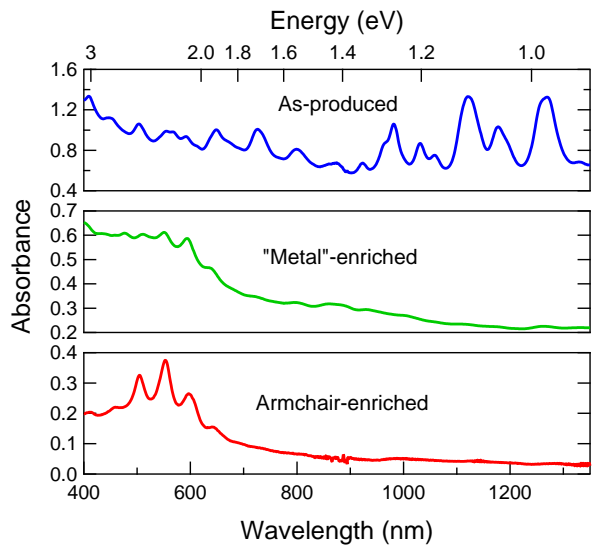}
\caption{\label{fig1-absorption}(color online) Optical absorption of the three nanotube suspensions studied: As-produced (AP-SWNT, blue), ``metal''-enriched (ME-SWNT, green), and armchair-enriched (AE-SWNT, red).  Traces are vertically offset for clarity.  Note ME-SWNT and AE-SWNT show significant suppression of features due to $\nu$~=~$\pm$1 SWNTs as compared to AP-SWNT, indicating the elimination of $\nu$~=~$\pm$1 species and hence $\nu$~=~0 enrichment by the DGU process.}
\end{figure}
%%%%%%%%%%%%%%%%%%%%

Figure 1 shows optical absorption spectra of the three HiPco SWNT samples.  The AP-SWNT sample (top, blue trace) displays all the typically observed features corresponding to the first ($E_{11}^{\rm S}$, 850-1600~nm) and second ($E_{22}^{\rm S}$, 570-850~nm) interband transitions of $\nu$~=~$\pm$1 tubes and the first ($E_{11}^{\rm M}$) interband transitions of $\nu$~=~0 tubes.  Based on absorption peak area estimates, this sample contains $\sim$40\% $\nu$~=~0 nanotubes~\cite{HarozetAl10ACS}, which agrees reasonably well with the expected value of 34\% from the number of possible ($n$,$m$) species contained within the diameter range (0.6-1.4~nm) of this particular material.  After DGU, however, a significant suppression of $\nu$~=~$\pm$1 features (650-1350~nm) is observed in both the ME-SWNT (middle, green trace) and AE-SWNT (bottom, red trace) samples with $\nu$~=~0 purity estimates around 90\% and 98\%, respectively.  While both DGU samples are strongly enriched in $\nu$~=~0 tubes, previous Raman studies based on radial breathing mode (RBM) intensities show that ME-SWNT samples are a {\em bulk} enrichment of all $\nu$~=~0 species~\cite{NiyogietAl09JACS} whereas AE-SWNT samples are chiral-angle-selective towards armchair ($n$ = $m$) chiralities~\cite{HarozetAl10ACS}.  This accounts for the sharper absorption features observed in the AE-SWNT sample due to the sizable reduction of the number of overlapping peaks from the various $\nu$~=~0 species.

%%%%% Figure 2 %%%%%
\begin{figure}
\includegraphics[scale=0.53]{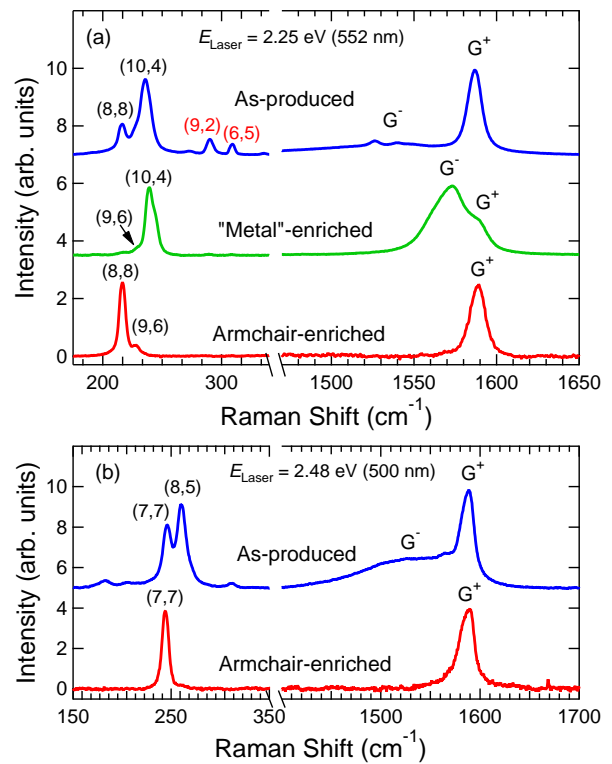}
\caption{\label{fig2-single line Raman}(color online) Raman spectra of RBM and G-band for AP-SWNT, ME-SWNT, and AE-SWNT samples taken at (a) 552~nm and (b) 500~nm where SWNTs from Families (2$n$+$m$)~=~24 and 21, respectively, are primarily probed.  In both sub-figures, appearance of the broad G$^{-}$ feature corresponds to resonance with non-armchair species such as (10,4) in ME-SWNT in (a) and (8,5) in AP-SWNT in (b).  In the case of sole resonance with armchair species [(8,8) and (7,7)], a single and narrow G$^{+}$ peak is observed.}
\end{figure}
%%%%%%%%%%%%%%%%%%%%

Figure 2(a) shows resonant Raman spectra for both the RBM and G-band ranges for all SWNT samples, collected at 552~nm laser excitation.  Comparing to existing experimental and theoretical Kataura plots in the literature, this wavelength primarily resonates with members of the (2$n$+$m$)~=~24 family [(8,8) and (10,4)] as well as some small-diameter $\nu$~=~1 species [(9,2) and (6,5)].  Examining the RBM region first, it is clearly confirmed that indeed the ME-SWNT (green trace) and AE-SWNT (red trace) samples are enriched in $\nu$~=~0 tubes while the AP-SWNT sample (blue trace) contains a mixture of both $\nu$~=~1 and $\nu$~=~0 species.  Furthermore, the ME-SWNT sample contains most of the members of Family 24, whereas the AE-SWNT sample contains only the (8,8) armchair species and a very small amount of the non-armchair species (9,6).

In the corresponding G-band region for the AP-SWNT (blue trace) in Fig.~2(a), we observe a G-band lineshape typical
of resonance with $\nu$~=~1 species such as (9,2) and (6,5),
%At first glance, this may appear surprising since the RBM intensities of the metallic species are significantly larger than that of the semiconductors present but can be explained when one takes a number of factors into consideration.  First, at this excitation wavelength neither the (9,2) or (6,5) are near their respective E$_{22}^S$ optical transitions ($\sim$543 and 568~nm~\cite{WeismanBachilo03NL}, respectively), which implies that the relative ratios of RBM intensities of each species at this wavelength is not necessarily representative of their relative abundances.  Only measurements taken at each ($n$,$m$) species' Raman energy maxima and corrected for the differences in ($n$,$m$)-dependent exciton-phonon coupling would allow an extraction of relative abundances.  Referring to the absorption spectrum of AP-SWNT in Fig.~1, we can see that there is in fact a large amount of (6,5) present in this sample by examining the absorbance of the $E_{11}^{\rm S}$ peak at 980~nm.  Despite this fact, we still observe a semiconducting-like G-band.  This is due to the fact that the Raman resonance window (considering resonance with both the incident and scattered photons) for the G-band ($\sim$100's meV) is substantially broader than that for the RBM ($\sim$100~meV), allowing species that may not strongly contribute to the RBM spectrum to contribute to the G-band spectrum.
because the intensity of the G$^{+}$ peak for $\nu$~=~$\pm$1 tubes is markedly stronger than that for $\nu$~=~0 tubes~\cite{MachonetAl06PRB,JiangetAl07PRB}.  With the removal of the $\nu$~=~$\pm$1 impurities, the ME-SWNT sample displays a G-band lineshape typically ascribed in the literature to the presence of only metallic nanotubes with the notably broad G$^{-}$ component [green trace in Fig.~2(a)].  However, the AE-SWNT sample, which is also free of $\nu$~=~$\pm$1 impurities, manifests a completely different G-band lineshape consisting of only a single and narrow G$^{+}$ component [red trace in Fig.~2(a)].  Such a large difference in G-band lineshape between these two samples suggests a significant ($n$,$m$)-dependence of the G$^{-}$ feature.  In fact, when taken together with the RBM spectrum for the ME-SWNT sample, a correlation between the appearance of small chiral-angle (or zigzag-like) $\nu$~=~0 species and that of the G$^{-}$ peak is evident.  Furthermore, this peak becomes absent at resonance with armchair tubes.

Examining another $\nu$~=~0 family, (2$n$+$m$)~=~21, we see a similar trend.  In Fig.~2(b), we present resonant Raman spectra of the RBM and G-band ranges for the AP-SWNT and AE-SWNT samples taken at 500~nm excitation.  At this wavelength, both samples are spectrally isolated from any $\nu$~=~$\pm$1 impurities, and, as such, only resonance with the $\nu$~=~0 tubes occurs.  AP-SWNT displays resonance with only two species, (8,5) and (7,7).  Correspondingly, we observe the broad G$^{-}$ feature and a narrow G$^{+}$ component.  In contrast, AE-SWNT resonates singly with the armchair species (7,7) and again displays a single, narrow G$^{+}$ peak as AE-SWNT in Fig.~2(a).

%Taken together, these results establish a trend of decreasing G$^{-}$ intensity with increasing resonance with large chiral-angle $\nu$~=~0 tubes to the point of complete extinction in the armchair limit.

%%%%% Figure 3 %%%%%
\begin{figure}
\includegraphics[scale=0.45]{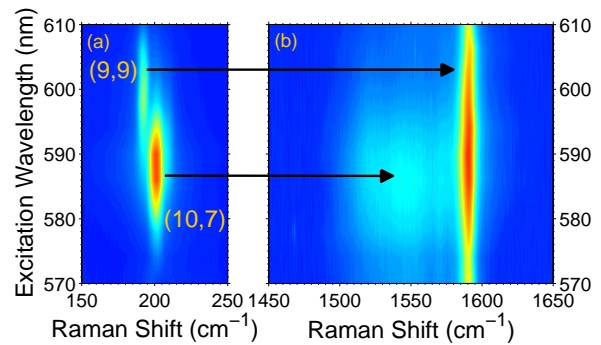}
\caption{\label{fig3-contour plot}(color online) Raman intensity for AE-SWNT taken as a function of Raman shift and excitation wavelength for Family (2$n$+$m$)~=~27.  (a) RBM region where two clear RBMs due to the (9,9) and (10,7) are observed.  (b) Corresponding G-band region where only the G$^{+}$ peak is observed when resonating primarily with the (9,9) and the appearance of the broad G$^{-}$ coincides with the maximum of the (10,7) RBM.}
\end{figure}
%%%%%%%%%%%%%%%%%%%%

To further examine the relationship between the appearance of the broad G$^{-}$ peak and ($n$,$m$) composition, we looked at Raman intensity of another isolated $\nu$~=~0 family, (2$n$+$m$)~=~27, as a function of both Raman shift and excitation wavelength.  Figure 3 shows a contour plot of Raman intensity of the AE-SWNT sample for the RBM and G-band regions over excitation wavelengths of 570-610~nm.  In Fig.~3(a), we see only two RBMs, from (9,9) and (10,7), again reflecting enrichment toward armchair-like species.  The corresponding G-band [Fig.~3(b)] shows only a single G$^+$ peak centered at $\sim$1590~cm$^{-1}$ at the longest excitation wavelengths ($\sim$610~nm) coinciding primarily with resonance of the (9,9) RBM.  As the wavelength is decreased, a broad G$^-$ peak appears in addition to the G$^+$ peak.  This broad G$^-$ peak, centered at $\sim$1550~cm$^{-1}$, reaches a maximum in intensity at $\sim$587~nm, which coincides with the RBM resonance maximum of the (10,7).  The simultaneous appearance of the G$^{-}$ peak with the Raman resonance maximum of the (10,7) and its absence with pure resonance with the (9,9) point out a clear correlation between ($n$,$m$) chirality and G-band lineshape.  Namely, the broad G$^{-}$ peak appears only in the presence of non-armchair $\nu$~=~0 nanotube species, and ($n$,$n$) armchair species consist of only one single, narrow peak.

%%%%% Figure 4 %%%%%
\begin{figure}
\includegraphics[scale=0.55]{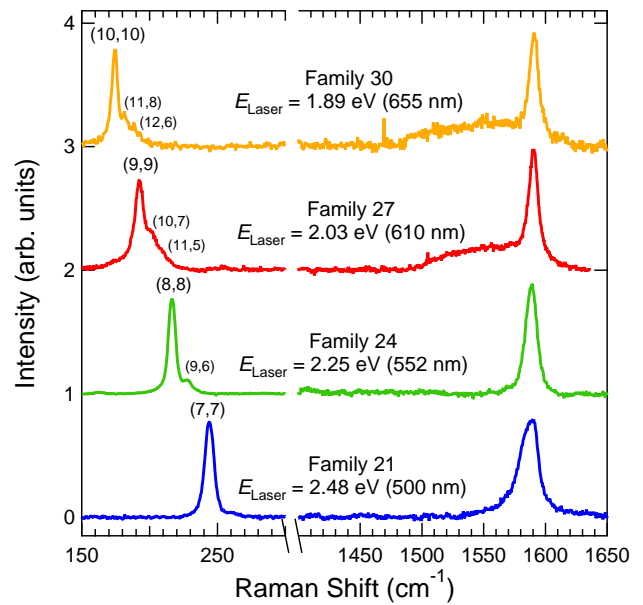}
\caption{\label{fig4-armchairs}(color online) Selected resonant Raman spectra for AE-SWNT taken at 655~nm, 610~nm, 552~nm, and 500~nm, where resonance primarily occurs with armchairs (10,10), (9,9), (8,8), and (7,7), respectively.  In each case, the G-band reflects contribution mainly from the G$^{+}$ peak only.}
\end{figure}
%%%%%%%%%%%%%%%%%%%%

Finally, Fig.~\ref{fig4-armchairs} shows selected Raman spectra of the (7,7), (8,8), (9,9), and (10,10) armchair families taken at 500, 552, 610, and 655~nm excitations, respectively, for the AE-SWNT sample.  These wavelengths are close to the general resonance maxima for each respective family.  Data taken across a diameter range of 0.96-1.38~nm shown here clearly demonstrates that the appearance and dominance of a single G$^{+}$ feature is a general result and indicator for the predominance of armchair species in a SWNT sample.  The observation of the weak G$^{-}$ peak at 610 and 655~nm represents the non-negligible but small contribution of residual non-armchair $\nu$~=~0 impurities [(10,7) and (11,8)] and a qualitative measure of the larger Raman cross-section such species have as compared to armchair tubes~\cite{JiangetAl07PRB}.

%%%%% Discussion %%%%%
Based on the aforementioned results, we attribute the narrow G$^{+}$ component observed during resonance with all $\nu$~=~0, armchair and non-armchair species to the transverse optical (TO) phonon mode and the broad G$^{-}$ component observed only with $\nu$~=~0, non-armchair species to the LO phonon mode arising from phonon softening due to Kohn anomaly~\cite{DubayetAl02PRL,LazzerietAl06PRB,IshikawaAndo06JPSJ,PiscanecetAl07PRB,NguyenetAl07PRL,WuetAl07PRL,FarhatetAl07PRL}.  The absence of the LO phonon feature for armchair tubes is consistent with some theoretical studies (e.g., \cite{DubayetAl02PRL,SaitoetAl01PRB}), i.e., it is {\em not} Raman active in armchairs, while it is inconsistent with more recent theories~\cite{SasakietAl08PRB,ParketAl09PRB}.  Furthermore, our results for G-band lineshape of armchair species agree with some single-tube Raman results~\cite{WuetAl07PRL,MicheletAl09PRB,BerciaudetAl10PRB} while disagreeing with others~\cite{ParketAl09PRB}.  The variability in single-tube Raman results comes from the sensitivity of G-band phonons to doping and charge transfer from intentional gating~\cite{NguyenetAl07PRL,WuetAl07PRL} and environmental effects that can induce a localization of electron density in the nanotube $\pi$~electron cloud~\cite{RaoetAl97Nature,ShimetAl08JPCC}.  As a result, generalizations of observed trends are difficult to establish at the single-tube level because only a few nanotubes are typically sampled in an experiment.  In contrast, the results presented here come from macroscopic ensembles containing $\sim$10$^{10}$ nanotubes per chirality.  Any tube-to-tube variations measured are essentially removed due to this large statistical sampling, and the resulting recorded spectrum represents the average response for each chirality.  In combination with the enrichment of armchairs and the resulting spectral isolation of each species, generalization of the absence of a broad LO peak to all armchair nanotubes is appropriate and justifiable.  Lastly, a refinement in the specificity of the use of the broad G$^{-}$ peak as a ``metallic'' signature can be made where its appearance is indicative of resonance with non-armchair, $\nu$~=~0 species.  However, the appearance of an only single, narrow G$^{+}$ signifies pure resonance with spectrally isolated armchair species.

%%%%% Conclusion %%%%%
In conclusion, using resonant Raman measurements of macroscopic ensembles of $\nu$~=~0 enriched SWNT samples, we have demonstrated that the appearance of a broad G$^{-}$ peak is due to the presence of non-armchair ``metallic'' tubes.  More importantly, the G-band of the truly gapless armchair tubes consists of a single, narrow G$^{+}$ component and serves as a diagnostic for their identification.  Because of the large statistical sampling presented here, these results are generalizable to all $\nu$~=~0 nanotubes. Finally, this study demonstrates the ability of using specialized, enriched samples such as these to extract meaningful and definitive information about single chiralities from a macroscopic scale, enabling future studies beyond the experimental limitations of the single-tube level.

%%%%% Acknowledgements %%%%%
\begin{acknowledgments}
This work was supported by the DOE/BES through Grant No.~DEFG02-06ER46308, the Robert A.~Welch Foundation through Grant No.~C-1509, the Air Force Research Laboratories under contract number FA8650-05-D-5807, and the LANL
LDRD Program.  This work was performed in part at the Center for Integrated Nanotechnologies, a U.S. Department of Energy, Office of Basic Energy Sciences user facility.  We thank C.~Kittrell, R.~H.~Hauge, and R.~Saito for useful discussions.
\end{acknowledgments}

%\bibliography{jun}

%%%%%%%%%%%%%%%%%%%%%%%%%%%%%%%%%%%%%%%%%%%%%%%%

\end{document}